\documentclass[fleqn,10pt,a4paper]{article}
\usepackage{amsmath,amssymb,bm}
\usepackage{cite,epsfig,color,graphicx}

\title{A kinetic model of radiating electrons} 

\author{A Noble\thanks{Department of Physics,
SUPA and University of Strathclyde, Glasgow, G4 0NG, UK}
\and
J Gratus\thanks{Department of Physics,
Lancaster University, Lancaster, LA1 4YB, UK
and Cockcroft Institute, Daresbury, WA4 4AD, UK}
\and
D A Burton\footnotemark[2]
\and
D A Jaroszynski\footnotemark[1]
}
\begin{document}
\maketitle
\begin{abstract}
A kinetic theory is developed to describe radiating electrons whose motion is governed by the Lorentz-Dirac equation. This gives rise to a generalized Vlasov equation coupled to an equation for the evolution of the physical submanifold of phase space. The pathological solutions of the $1$-particle theory may be removed by expanding the latter equation in powers of $\tau := q^2/ 6\pi m$. The radiation-induced change in entropy is explored, and its physical origin is discussed. As a simple demonstration of the theory, the radiative damping rate of longitudinal plasma waves is calculated.
\end{abstract}
\newcommand{\xd}{\dot{x}}
\newcommand{\qm}{\frac{q}{m}}
\newcommand{\xdd}{\ddot{x}}
\newcommand{\xddd}{\dddot{x}}
\section{Introduction}
The question of how a charged particle interacts with its own radiation field remains unclear, despite intensive theoretical investigations over the last century.  Until recently, this issue has been largely of theoretical interest as in current facilities the self-force is small in comparison to the Lorentz force due to the applied fields. However, the advent of new ultra-high intensity laser facilities requires that radiation reaction effects be taken seriously.  For instance, the Extreme Light Infrastructure (ELI)~\cite{ELI} is expected to operate with intensities exceeding $10^{23}$ W cm$^{-2}$, and electron energies in the GeV range, at which level the radiation reaction force becomes comparable to and can even exceed the Lorentz force.

The problem was first addressed by Lorentz~\cite{Lorentz} and Abraham~\cite{Abraham}, who introduced a third order differential equation to describe the trajectory of a radiating particle. This work was later generalized to the relativistic regime by Dirac~\cite{Dirac}, leading to what is now known as the Abraham-Lorentz-Dirac equation or, more simply, the Lorentz-Dirac equation,
\begin{equation}
\ddot{C}^a= -\qm F^a{}_b\dot{C}^b + \tau \Delta^a{}_b\dddot{C}^b,  \label{LD}
\end{equation}
for the components $\{C^a\}$ of the worldline $C: \lambda \mapsto x^a = C^a(\lambda)$ of an electron of charge $q$ and mass $m$ in an electromagnetic field $F_{ab}$ expressed in inertial coordinates $(x^a)$ on Minkowski spacetime $\cal M$ with metric tensor $\eta$,
\begin{equation}
\eta= -dx^0 \otimes dx^0+dx^1 \otimes dx^1+dx^2 \otimes dx^2+dx^3 \otimes dx^3.
\end{equation}
 $\tau := q^2/6\pi m\simeq 10^{-23}$s is the characteristic time of the electron\footnote{For definiteness we consider electrons, though the formalism could be applied to other charged particles.}, $\Delta^a{}_b:= \delta^a_b+\dot{C}^a \dot{C}_b$ is the $\dot{C}$-orthogonal projection, and an overdot indicates differentiation with respect to proper time $\lambda$,
\begin{equation}
\label{normalization}
\eta_{ab} \dot{C}^a \dot{C}^b = -1 ,
\end{equation}
where $[\eta_{ab}]={\rm diag}(-1,1,1,1)$. We work in Heaviside-Lorentz units with $c=1$, and raise and lower indices with $\eta_{ab}$.  Latin indices run from 0 to 3, and the Einstein summation convention is used.

Equation (\ref{LD}) has proved highly controversial.  The difficulties stem from the fact that it contains derivatives of the acceleration, so specifying the initial position and velocity is not sufficient to determine the solution.  On the other hand, a generic specification of the initial acceleration leads to exponentially growing proper acceleration,
\begin{equation}
\sqrt{\ddot{C}^a\ddot{C}_a}\sim e^{\lambda/\tau},
\end{equation}
even in the absence of applied forces.  Such `runaway solutions' are clearly at odds with our observations, and so must be eliminated.

The solution proposed by Dirac was to use the elimination of the nonphysical solutions, rather than the initial acceleration, as part of the initial data, effectively converting the Lorentz-Dirac equation into a second order integro-differential equation, 
\begin{align}
\nonumber
&\ddot{C}^a(\lambda)= \int^\infty_0 K^a(\lambda+\alpha\tau)e^{-\alpha}d\alpha, \\
\label{LD_integro-differential}
&K^a:=-\qm F^a{}_b \dot{C}^b-\tau \ddot{C}^b\ddot{C}_b \dot{C}^a.
\end{align}
Equation (\ref{LD_integro-differential}) requires the acceleration at a given time to depend on the force applied at all subsequent times, and so is acausal.  However, the acausal effects are exponentially damped for times $\gtrsim \tau$ into the future, and so in practice are unobservable.

Despite the many distinct derivations of the Lorentz-Dirac equation \cite{Dirac,Bhabha,Feynman,Rohrlich1,Teitelboim,Barut,Gratus,Ferris}, the problems of runaway solutions and acausality have led many researchers to propose alternative descriptions of radiation reaction \cite{Eliezer, Ford, Sokolov, Hammond}.  The most widely adopted of these is the Landau-Lifshitz equation \cite{Landau},
\begin{equation}
\ddot{C}^a= -\qm F^a{}_b \dot{C}^b - \qm \tau \big[ \partial_d F^a{}_b \dot{C}^b- \qm \Delta^a{}_b F^b{}_c F^c{}_d\big] \dot{C}^d,  \label{LL}
\end{equation}
obtained by perturbing (\ref{LD}) about the Lorentz force and neglecting terms higher than first order in $\tau$.  It has been argued by Spohn \cite{Spohn} that the solutions approximated by (\ref{LL}) are precisely those corresponding to Dirac's asymptotic condition (\ref{LD_integro-differential}).  Further, it is often claimed \cite{Rohrlich} that effects neglected in the derivation of the LD equation -- such as spin and quantum effects -- are of order $\tau^2$, so that (\ref{LL}) is as accurate as (\ref{LD}).  Though plausible in the case of a single radiating electron, this claim remains unproven, and it is more reasonable to consider (\ref{LL}) an approximation to (\ref{LD}). There has been some interest recently in exploring the validity of this approximation \cite{Griffiths,Bulanov}.

In practice, radiation reaction is unlikely to ever be observed in the context of a single radiating particle. However, modern laser facilities accelerate electron bunches with charge of the order of $10\,{\rm pC}$, containing $10^8$ particles, and it follows that an appropriate description of radiation reaction is not the one-particle equation of motion, but rather a kinetic theory. A kinetic theory based on (\ref{LL}) has appeared in the literature~\cite{Tamburini, Berezhiani, Hazeltine}. However, it is not clear {\it a priori} that the approximations leading to (\ref{LL}) are appropriate for describing a large number of interacting electrons.

The Vlasov equation describes a large collection of particles as a continuum. Charged continua typically do not suffer from the pathologies associated with radiating point particles, and since the Maxwell-Vlasov system satisfies the energy and momentum conservation laws that led to (\ref{LD}) it may seem that it does not need modifying to describe radiation reaction. However, the continuum comprises `particles' with infinitesimal mass and charge, $(q,m) \rightarrow 0$ with $q/m$ fixed, in which limit (\ref{LD}) becomes the usual Lorentz force equation. To describe a collection of real electrons with finite $q$ and $m$, the Vlasov equation must be modified so that elements of the continuum follow trajectories of the Lorentz-Dirac equation (\ref{LD}). Such a kinetic equation has been introduced in \cite{Hakim}; however, in that work the physical phase space was not identified, so the 1-particle distribution is required to be distributional (in the sense of Schwartz). As well as impeding the physical interpretation, this has led to a number of errors, for example in calculating the rate of change of entropy.

In this article, we introduce a kinetic theory for electrons obeying (\ref{LD}), with the 1-particle distribution a regular function on the physical phase space, and explore some of its consequences.

\section{Kinetic theory with radiation reaction}

The usual approach to obtaining an equation for a large collection of particles from the equation of motion for a single particle is to consider a $1$-particle distribution on a subspace of $T\cal M$.  This distribution is then taken to be constant along the lifts of particle orbits to $T\cal M$, which may be obtained as integral curves of a vector field on $T\cal M$. This prescription is not quite general enough to incorporate radiation reaction, since the Lorentz-Dirac equation is third order. The approach adopted in the present article is to consider a 1-particle distribution on a subspace of the bundle with total space
\begin{equation}
T {\cal M} \oplus T {\cal M}= \bigcup_{p\in\cal M} T_p {\cal M}\oplus T_p {\cal M}
\end{equation}
and whose fibres are two copies of the tangent space, one representing $4$-velocity and the other representing $4$-acceleration, with $(\xd^a)$ and $(\xdd^a)$ the induced coordinates on the first and second copies of the tangent space. We are interested in the physical subspace ${\cal Q} \subset T{\cal M}\oplus T{\cal M}$ given by 
\begin{equation}
{\cal Q}= \{ (x,\xd,\xdd) \in T{\cal M}\oplus T{\cal M} | \ \varphi_1=0,\ \varphi_2=0, \dot{x}^0 > 0\}
\end{equation}
where
\begin{align}
&\varphi_1= \frac{1}{2}(\eta_{ab} \xd^a \xd^b +1),\\
&\varphi_2= \eta_{ab}\xd^a \xdd^b.
\end{align}
Worldlines parametrized by proper time satisfy $\varphi_1=0$, from which it follows that their $4$-velocity and $4$-acceleration should be orthogonal as encoded in $\varphi_2=0$. It is straightforward to show that solutions to the Lorentz-Dirac equation are integral curves of the vector field $L$,
\begin{equation}
\label{LD_vector_field}
L := \xd^a \frac{\partial}{\partial x^a}+ \xdd^a \frac{\partial}{\partial \xd^a}+ \big[\xdd^b \xdd_b \xd^a+ \tau^{-1} (\xdd^a+ \qm F^a{}_b\xd^b)\big] \frac{\partial}{\partial \xdd^a},
\end{equation}
on $T{\cal M}\oplus T{\cal M}$.
 
Since $L$ is tangent to $\cal Q$ ($L\varphi_1=\varphi_2$ and $L\varphi_2=2\eta_{ab}\xdd^a\xdd^b\varphi_1+ \tau^{-1}\varphi_2$) it can be expressed as the push-forward of a vector field $L_{\cal Q}$ on $\cal Q$, $L=\iota_* L_{\cal Q}$, where $\iota: {\cal Q} \hookrightarrow T{\cal M}\oplus T{\cal M}$ is the inclusion map. The Liouville vector field $L_{\cal Q}$ is calculated explicitly below.

The $1$-particle distribution must satisfy the Vlasov equation, which states that the $1$-particle distribution is preserved under the flow of the Liouville vector field. However, it is important to note that it is not the particle density $f$ which must be preserved along the flow, but rather the particle distribution $f\omega$, where $\omega$ is a non-vanishing top-dimensional form. Thus, the Vlasov equation may be cast as
\begin{equation}
\label{VlasQ_actual}
{\cal L}_{L_{\cal Q}} (f\omega) = 0
\end{equation}
where ${\cal L}_X$ is the Lie derivative with respect to $X$, $f$ is a $0$-form on ${\cal Q}$, and $\omega$ is a non-vanishing $10$-form on ${\cal Q}$.

In principle, $\omega$ may be {\it any} measure on $\cal Q$; however, in practice, it is convenient to use the Leray measure of ${\cal Q} \subset T{\cal M}\oplus T{\cal M}$ derived from the natural measure $\hat{\omega}$ on $T{\cal M}\oplus T{\cal M}$,
\begin{equation}
\hat{\omega}= dx^{0123}\wedge d\xd^{0123}\wedge d\xdd^{0123},
\end{equation}
where $dx^{0123}\equiv dx^0\wedge dx^1\wedge dx^2\wedge dx^3$, etc.  The Leray measure $\omega$ is defined by
\begin{equation}
\omega := \iota^* \tilde{\omega},
\end{equation}
where $\tilde{\omega}$ is any 10-form on $T{\cal M}\oplus T{\cal M}$ satisfying
\begin{equation}
\label{hat_omega}
\hat{\omega}= \tilde{\omega}\wedge d\varphi_1 \wedge d\varphi_2.
\end{equation}
We choose
\begin{equation}
\label{tilde_omega}
\tilde{\omega}= (\xd^1 \xdd^0-\xd^0\xdd^1)^{-1} dx^{0123}\wedge d\xdd^{0123}\wedge d\xd^{23},
\end{equation}
which can readily be shown to satisfy (\ref{hat_omega}).

To calculate $\iota^*\tilde{\omega}$, we use coordinates ($x^a, v^\mu, a^\nu$) on $\cal Q$ such that
\begin{align}
\label{Q_coords_1}
&\iota^\ast x^a = x^a,\\
\label{Q_coords_2}
&\iota^\ast \xd^0= \sqrt{1+{\boldsymbol v}^2}, \qquad\qquad \iota^\ast \xd^\mu= v^\mu,\\
\label{Q_coords_3}
&\iota^\ast \xdd^0=\frac{{\boldsymbol a}\cdot{\boldsymbol v}}{\sqrt{1+{\boldsymbol v}^2}}, \qquad\qquad  \iota^\ast \xdd^\mu= a^\mu,
\end{align}
where Greek indices run from $1$ to $3$ and ${\boldsymbol a}\cdot {\boldsymbol v} \equiv a^\mu v_\mu$, etc. 

Using (\ref{tilde_omega}), (\ref{Q_coords_1}), (\ref{Q_coords_2}), (\ref{Q_coords_3}) it follows that the Leray measure $\omega= \iota^* \tilde{\omega}$ on ${\cal Q}$ is
\begin{equation}
\omega = \frac{1}{1+{\boldsymbol v}^2} dx^{0123}\wedge da^{123}\wedge dv^{123}.
\end{equation}
Similarly, on the fibre ${\cal Q}_p = \pi^{-1}(p)$ of the bundle $({\cal Q},\pi,{\cal M})$ over $p\in \cal M$, the induced measure $\Omega$ is identified as
\begin{equation}
\Omega := \frac{da^{123}\wedge dv^{123}}{1+{\boldsymbol v}^2} 
\end{equation}
because we adopt $dx^{0123}$ as the measure on ${\cal M}$\footnote{The same procedure leads to $dx^{0123}\wedge dv^{123}/\sqrt{1+{\boldsymbol v}^2}$ for the Leray measure on the bundle with total space ${\cal E} := \{ (x,\xd) \in T{\cal M} | \ \varphi_1 =0 , \dot{x}^0 > 0\}$, where $v^\mu$ is the pull-back of $\dot{x}^\mu$ to ${\cal E}$ with respect to the inclusion ${\cal E} \hookrightarrow T{\cal M}$. It follows that the induced measure on a fibre of that bundle is $dv^{123}/\sqrt{1+{\boldsymbol v}^2}$ as expected.}.


It may be shown that
\begin{align}
&\iota_* \frac{\partial}{\partial x^a} = \frac{\partial}{\partial x^a},\\
&\iota_* \frac{\partial}{\partial v^\mu}= \frac{\partial}{\partial \xd^\mu}+ \frac{v_\mu}{\sqrt{1+{\boldsymbol v}^2}} \frac{\partial}{\partial \xd^0}+ \frac{a_\nu}{\sqrt{1+{\boldsymbol v}^2}}\bigg( \delta^\nu_\mu- \frac{v^\nu v_\mu}{1+{\boldsymbol v}^2} \bigg) \frac{\partial}{\partial \xdd^0},\\
&\iota_* \frac{\partial}{\partial a^\mu}= \frac{\partial}{\partial \xdd^\mu}+ \frac{v_\mu}{\sqrt{1+{\boldsymbol v}^2}} \frac{\partial}{\partial \xdd^0}
\end{align}
and it follows $\iota_* L_{\cal Q} = L$ where
\begin{equation}
\label{LQ}
L_{\cal Q}= \xd^a \frac{\partial}{\partial x^a}+ a^\mu \frac{\partial}{\partial v^\mu}+ \bigg( \xdd^a\xdd_a v^\mu+ \tau^{-1} (a^\mu+ \frac{q}{m}F^\mu{}_a \xd^a) \bigg) \frac{\partial}{\partial a^\mu}.
\end{equation}
In (\ref{LQ}), and from now on, it is to be understood that $\xd^a$ and $\xdd^a$ are shorthand for the values of $\iota^* \xd^a$ and $\iota^* \xdd^a$ respectively.

In the absence of radiation reaction, the Leray measure on the total space $\cal E$ is preserved by the flow of the Liouville vector field induced from the Lorentz force equation. This well-known result is commonly described as ``conservation of phase space volume''. By contrast, the Leray measure $\omega$ on $\cal Q$ is not preserved by the flow of $L_{\cal Q}$:
\begin{equation}
\label{omega_evolution}
{\cal L}_{L_{\cal Q}}\omega = \frac{3}{\tau}\omega.
\end{equation}
This may be understood physically as a consequence of losses due to radiation. It follows from (\ref{omega_evolution}) that (\ref{VlasQ_actual}) can be written
\begin{equation}
L_{\cal Q} f+ \frac{3}{\tau}f=0.  \label{VlasQ}
\end{equation}

The electrons couple to the electromagnetic field through the final term in (\ref{LQ}), and through the Maxwell equations
\begin{align}
\nonumber &\frac{\partial F_{bc}}{\partial x^a}+\frac{\partial F_{ca}}{\partial x^b}+\frac{\partial F_{ab}}{\partial x^c}=0,\\
 &\frac{\partial F^{ab}}{\partial x^a}= J^b+ J^b_{\rm ext},  \label{Max}
\end{align}
where the electron current is given by
\begin{equation}
J^a= q\int f\xd^a \Omega  \label{currentQ}
\end{equation}
and $J^a_{\rm ext}$ is the current of any source other than the electrons.

\section{Physical solutions}

Equation (\ref{VlasQ}) does not in general reduce to the usual Vlasov equation in the limit $\tau\rightarrow 0$.  This is because, for a regular solution to (\ref{VlasQ}), $f$ will be nonzero over a range of $\boldsymbol a$ for given values of $(x, \boldsymbol v)$, and so will necessarily include runaway solutions as well as physical ones.  To avoid this, look for solutions of the form
\begin{equation}
f(x,{\boldsymbol v,\boldsymbol a})= \sqrt{1+{\boldsymbol v}^2}g(x,{\boldsymbol v})\delta^{(3)}\big({\boldsymbol a}-{\boldsymbol A}(x,{\boldsymbol v})\big),  \label{Schwartz}
\end{equation}
where the (as yet undetermined) functions $A^\mu$ describe the submanifold of phase space containing physical trajectories.  The factor $\sqrt{1+{\boldsymbol v}^2}$ in (\ref{Schwartz}) ensures that $g$ can be interpreted as a particle distribution function on the space of unit timelike $4$-vectors, from which it follows that the current (\ref{currentQ}) becomes
\begin{equation}
J^a= q\int g\xd^a \frac{d^3v}{\sqrt{1+{\boldsymbol v}^2}}.
\end{equation}

Using (\ref{Schwartz}) in (\ref{VlasQ}) and integrating out the acceleration variables leads to the coupled equations
\begin{align}
& \xd^a \frac{\partial A^\mu}{\partial x^a}+ A^\nu \frac{\partial A^\mu}{\partial v^\nu}= A^aA_a v^\mu+ \frac{1}{\tau} (A^\mu+\qm F^\mu{}_a \xd^a),  \label{Accel}\\
& \xd^a \frac{\partial g}{\partial x^a}+ \sqrt{1+{\boldsymbol v}^2}\frac{\partial}{\partial v^\mu} \Big( \frac{gA^\mu}{\sqrt{1+{\boldsymbol v}^2}}\Big)=0,  \label{Vlasg}
\end{align}
with $A^0:=v^\mu A_\mu/\sqrt{1+{\boldsymbol v}^2}$. Equation (\ref{Accel}) describes the evolution of the physical submanifold, as governed by (\ref{LD}), while (\ref{Vlasg}) is a generalized Vlasov equation for the 1-particle distribution $g$.  
It is the reduced system (\ref{Accel})--(\ref{Vlasg}), rather than (\ref{VlasQ}), that reduces to the usual Vlasov equation in the limit $\tau\rightarrow 0$.

So far we have placed no restrictions on the submanifold represented by $A^\mu$, other than that it satisfy (\ref{Accel}).  In order that it represent the physical solutions to (\ref{LD}), it must be regular in the limit $\tau \rightarrow 0$.  $A^\mu$ may then be expanded in powers of $\tau$:
\begin{equation}
A^\mu= \sum^\infty_{n=0} \tau^n A^\mu_{(n)},
\end{equation}
with
\begin{align}
& A^\mu_{(0)}= -\qm F^\mu{}_{a(0)} \xd^a,\\
& A^\mu_{(n+1)}= -\qm F^\mu{}_{a(n+1)}\xd^a+ \xd^a\frac{\partial A^\mu_{(n)}}{\partial x^a}
+ \sum^n_{j=0} A^\nu_{(n-j)} \frac{\partial A^\mu_{(j)}}{\partial v^\nu}- v^\mu \sum^n_{j=0} A^a_{(n-j)} A_{a(j)}.
\end{align}
Although $\tau$ appears explicitly only in the acceleration equation (\ref{Accel}), the distribution $g$ and electromagnetic field $F_{ab}$ acquire a $\tau$ dependence via their couplings to $A^\mu$ in the Vlasov equation (\ref{Vlasg}) and the Maxwell equations (\ref{Max}), respectively, and may be similarly expanded:
\begin{equation}
g = \sum^\infty_{n=0} \tau^n g_{(n)}, \qquad  F^{ab}= \sum^\infty_{n=0} \tau^n F^{ab}_{(n)}.
\end{equation}

Truncating at $n=0$ yields the usual Vlasov equation without radiation reaction, while truncating at $n=1$ yields the kinetic theory derived from the Landau-Lifshitz equation (\ref{LL}).  This supports the large body of work \cite{Tamburini, Berezhiani, Hazeltine} utilizing the latter approach.


\section{Entropy}

In the absence of radiation reaction, the entropy of the electrons is conserved.  This is not generally true if radiation reaction is included, as the velocity-divergence of the physical acceleration acts as a source of entropy.  

Consider the entropy density $s=-g\ln g$ on phase space, which from the Vlasov equation (\ref{Vlasg}) satisfies
\begin{equation}
 \xd^a \frac{\partial s}{\partial x^a}+ \sqrt{1+{\boldsymbol v}^2}\frac{\partial}{\partial v^\mu}\bigg(\frac{sA^\mu}{\sqrt{1+{\boldsymbol v}^2}}\bigg)
 = g\sqrt{1+{\boldsymbol v}^2}\frac{\partial}{\partial v^\mu}\bigg( \frac{A^\mu}{\sqrt{1+{\boldsymbol v}^2}}\bigg).
\end{equation}
Then the entropy current $s^a= \int \xd^a s d^3 v/\sqrt{1+{\boldsymbol v}^2}$ on spacetime satisfies
\begin{equation}
\frac{\partial s^a}{\partial x^a}= \int g\frac{\partial}{\partial v^\mu} \bigg(\frac{A^\mu}{\sqrt{1+{\boldsymbol v}^2}}\bigg)d^3v,
\end{equation}
and the change in the total entropy $S=\int s^0 dx$ of the electrons is
\begin{equation}
\frac{dS}{dt}= \int g\frac{\partial}{\partial v^\mu} \bigg(\frac{A^\mu}{\sqrt{1+{\boldsymbol v}^2}}\bigg)d^3v d^3x. \label{entrate}
\end{equation}
Although the entropy density $s$ is defined with respect to a reference density, conservation of particle number ensures that changing this reference density merely shifts the total entropy by a constant, so the rate of change of entropy (\ref{entrate}) is well-defined.

Restricting to the physical solutions of (\ref{Accel}), we can expand $S= \sum^\infty_{n=0} \tau^n S_{(n)}$, and it follows that
\begin{equation}
\frac{dS_{(n)}}{dt}= \sum^n_{j=0} \int g_{(n-j)}\frac{\partial}{\partial v^\mu} \bigg(\frac{A^\mu_{(j)}}{\sqrt{1+{\boldsymbol v}^2}}\bigg)d^3v d^3x.
\end{equation}

The contribution from $A^\mu_{(0)}$ vanishes, so the leading order entropy change comes from $n=1$,
\begin{equation}
\frac{dS_{(1)}}{dt}= - \frac{1}{m}\int \Big( J_a (J^a+J^a_{\rm ext})+ 4 \frac{q^2}{m^2} T_{ab} S^{ab} \Big) d^3 x,  \label{entchange}
\end{equation}
where 
\begin{equation}
T_{ab}= F_{ac} F_b{}^c- \frac{1}{4} \eta_{ab}F_{cd}F^{cd}
\end{equation}
is the stress-energy-momentum tensor of the electromagnetic field and 
\begin{equation}
S^{ab}= m\int g \xd^a \xd^b \frac{d^3v}{\sqrt{1+{\boldsymbol v}^2}}
\end{equation}
is that of the electrons.  Note that $J^a$ and $S^{ab}$ in (\ref{entchange}) are calculated in the limit $\tau\rightarrow 0$.

The rate of change of entropy (\ref{entchange}) has previously been derived in \cite{Tamburini} from a kinetic theory based on the Landau-Lifshitz equation, though the physical interpretation was there less apparent.  The self-interaction term, $J_a J^a$, increases the entropy, while the interaction with the field, $T_{ab}S^{ab}$, leads to an entropy decrease.  The effect of the remaining term, $J_aJ^a_{\rm ext}$, depends on the nature of any external currents. In \cite{Hakim}, only the $J_aJ^a$ contribution was found, leading to the incorrect statement that the entropy of the electrons always increases. In fact, it is quite possible for the $T_{ab}S^{ab}$ term to dominate, leading to radiation cooling of the electrons.


\section{Plasma oscillations}

A simple demonstration of the theory comes from exploring the effect of radiation reaction on plasma waves.  Linearize (\ref{Max}), (\ref{Accel})--(\ref{Vlasg}) about the background
\begin{equation}
g= \widehat{g}({\boldsymbol v}), \qquad A^\mu=0, \qquad F_{ab}=0,
\end{equation}
with $J^a_{\rm ext}=-qn_0 \delta^a_0$ and $n_0=\int \widehat{g} d^3v$.  Assuming longitudinal perturbations of the form $\exp i(kz-\omega t)$ yields the dispersion relation
\begin{equation}
1= \frac{q^2}{m} \int \frac{\big( 1+v^2_1+v^2_2\big)\widehat{g}}{\Delta(\omega\sqrt{1+{\boldsymbol v}^2}-kv_3)^2}\frac{d^3v}{\sqrt{1+{\boldsymbol v}^2}},  \label{disp}
\end{equation}
where $\Delta=  1+i\tau(\omega\sqrt{1+{\boldsymbol v}^2}-kv_3)$ represents the modification of the plasma waves due to radiation reaction.

The dispersion relation (\ref{disp}) possesses a more complicated mode structure than its counterpart in the Maxwell-Vlasov system.  As well as small modifications to the usual solutions, there exist entirely new roots.  However, the latter do not exist in the limit $\tau\rightarrow 0$, and so must be rejected as unphysical.

Taking the cold equilibrium $\widehat{g}({\boldsymbol v})=n_0\delta^{(3)}({\boldsymbol v})$ yields
\begin{equation}
\omega^2_p= \omega^2(1+i\tau \omega),  \label{colddisp}
\end{equation}
where $\omega_p=\sqrt{q^2n_0/m}$ is the plasma frequency.  Though rather simplistic, this nicely illustrates the effects of incorporating radiation reaction into the Vlasov equation.  Two (physically equivalent) roots of (\ref{colddisp}),
\begin{equation}
\omega \approx \pm \omega_p \big(1-\frac{5}{8}(\omega_p\tau)^2\big)- \frac{i}{2}\omega^2_p \tau,  \label{physdamp}
\end{equation}
indicate a damping rate of order $\tau$ and an order $\tau^2$ frequency downshift.  There also exists a third (purely imaginary) root,
\begin{equation}
\omega \approx \frac{i}{\tau},  \label{nonphys}
\end{equation}
which represents extremely rapid growth, without oscillation.  This mode, which does not exist in the limit $\tau\rightarrow 0$, corresponds to the runaway solutions of the Lorentz-Dirac equation, and must be discarded as nonphysical.  Results equivalent to (\ref{physdamp}) may be obtained by truncating the expansion in $\tau$ at $n=2$, while the root (\ref{nonphys}) is excluded in this approach.  Similar results may be found for other types of waves in plasmas.


\section{Conclusions}
In summary, with the advent of ultra-high intensity laser facilities, such as ELI, it is important to have a reliable kinetic theory of radiating particles, incorporating radiation reaction.  We have developed such a theory based on the full Lorentz-Dirac equation, and found that it reduces to the usual Vlasov theory and to a kinetic theory based on the Landau-Lifshitz equation in appropriate limits.  As simple demonstrations of the theory, we have explored the effects of radiation reaction on entropy and on longitudinal plasma waves.

\section{Acknowledgments}
We would like to thank Robin Tucker for useful discussions.  This work was supported by UK EPSRC, the Laserlab-Europe consortium and the FP7--Extreme Light Infrastructure (ELI) project.

\end{document}